\documentclass[twoside,twocolumn,9pt]{article}
\usepackage{extsizes}
\usepackage[super,sort&compress,comma]{natbib} 
\usepackage[version=3]{mhchem}
\usepackage[left=1.5cm, right=1.5cm, top=1.785cm, bottom=2.0cm]{geometry}
\usepackage{balance}
\usepackage{widetext}
\usepackage{times,mathptmx}
\usepackage{sectsty}
\usepackage{graphicx} 
\usepackage{lastpage}
\usepackage[format=plain,justification=raggedright,singlelinecheck=false,font={stretch=1.125,small,sf},labelfont=bf,labelsep=space]{caption}
\usepackage{float}
\usepackage{fancyhdr}
\usepackage{fnpos}
\usepackage[english]{babel}
\usepackage{array}
\usepackage{droidsans}
\usepackage{charter}
\usepackage[T1]{fontenc}
\usepackage[usenames,dvipsnames]{xcolor}
\usepackage{setspace}
\usepackage[compact]{titlesec}

\usepackage{subcaption}

\usepackage{color}

\usepackage{epstopdf}

\definecolor{cream}{RGB}{222,217,201}

\begin{document}

\pagestyle{fancy}
\thispagestyle{plain}
\fancypagestyle{plain}{

}




\fancyhead{}
\renewcommand{\headrulewidth}{0pt} 

\makeatletter 
\newlength{\figrulesep} 
\setlength{\figrulesep}{0.5\textfloatsep} 

\newcommand{\topfigrule}{\vspace*{-1pt}%
\noindent{\color{cream}\rule[-\figrulesep]{\columnwidth}{1.5pt}} }

\newcommand{\botfigrule}{\vspace*{-2pt}%
\noindent{\color{cream}\rule[\figrulesep]{\columnwidth}{1.5pt}} }

\newcommand{\dblfigrule}{\vspace*{-1pt}%
\noindent{\color{cream}\rule[-\figrulesep]{\textwidth}{1.5pt}} }

\makeatother

\twocolumn[
  \begin{@twocolumnfalse}
\sffamily
\begin{tabular}{m{2.cm} p{13.5cm} }

 & \noindent\LARGE{\textbf{Domain size polydispersity effects on the structural and dynamical properties in lipid monolayers with phase coexistence.
}} \\

\vspace{0.3cm} & \vspace{0.3cm} \\

 & \noindent\large{Elena Rufeil Fiori$^{1,2,\ast}$ and Adolfo J. Banchio$^{1,2}$} \\

 & \noindent\normalsize{

Lipid monolayers with phase coexistence are a frequently used model for lipid membranes. In these systems, domains of the liquid--condensed phase always present size polydispersity. 
However, very few theoretical works consider size distribution effects on the monolayer properties.
Because of the difference in surface densities, domains have excess dipolar density with respect to the surrounding liquid expanded phase, originating a dipolar inter--domain interaction. 
This interaction depends on the domain area, and hence the presence of a domain size distribution is associated with interaction polydispersity.
Inter--domain interactions are fundamental to understanding the structure and dynamics of the monolayer. 
For this reason, it is expected that  polydispersity significantly alters  monolayer properties. 
By means of Brownian dynamics simulations, we study the radial distribution function (RDF), the average mean square displacement and the average time--dependent self--diffusion coefficient, $D(t)$, of lipid monolayers with normal distributed size domains. 
For this purpose, we vary the relevant system parameters, polydispersity and interaction strength, within a range of experimental interest.
We also analyze the consequence of using a monodisperse model for determining the interaction strength from an experimental RDF.
It was found that polydispersity strongly affects the value of the interaction strength obtained, which is greatly underestimated if polydispersity is not considered. 
However, within a certain range of parameters, the RDF obtained from a polydisperse model can be well approximated by that of a monodisperse model, suitably fitting the interaction strength, even for $40\%$ polydispersities.
For small interaction strengths or small polydispersities, the polydisperse systems obtained from fitting the experimental RDF have an average mean square displacement and $D(t)$ in good agreement with that of the monodisperse system.

} \\

\end{tabular}

 \end{@twocolumnfalse} \vspace{0.6cm}

  ]

\renewcommand*\rmdefault{bch}\normalfont\upshape
\rmfamily
\section*{}
\vspace{-1cm}


\footnotetext{\textit{$^1$ Universidad Nacional de C\'ordoba, Facultad de Matem\'atica, Astronom\'ia, F\'isica y Computaci\'on, C\'ordoba, Argentina.}}
\footnotetext{\textit{$^2$ Consejo Nacional de Investigaciones Cient\'ificas y T\'ecnicas, CONICET, IFEG, C\'ordoba, Argentina.}}
\footnotetext{\textit{$\ast$~e-mail: rufeil@famaf.unc.edu.ar}}



\section{Introduction}

Most biologically relevant monolayers
present phase coexistence characterized by domains formed by lipids in an ordered phase state, dispersed in a continuous, disordered phase~\cite{duncan11, dhar12, baoukina16, min17}.
The domains interact with each other~\cite{andelman85, andelman86,ursell09}, and these interactions affect their own movement~\cite{wilke14,wilke10} as well as that of other species present in the monolayer~\cite{forstner08,ruckerl08}.
Inter--domain interaction may be related to electrostatic forces (dipolar or Coulombic repulsions), forces related to the spontaneous curvature of the coexisting phases and hydrodynamic forces that appear when domains are in motion.

Dipolar inter--domain interaction, which is always present, arises from the excess dipolar density of the ordered phase with respect to the continuous phase. Hence, the dipolar strength is proportional to the domain area. 

Lipid monolayers typically exhibit domain size polydispersity
~\cite{mulder03, hu06, lee11, wilke17}.
In particular, Langmuir monolayers at the air--water interface show a wide domain size distribution~\cite{forstner08, wilke10, rufeil16, lee11}.
Due to the origin of the dipolar interaction, size polydispersity leads to interaction polydispersity, which usually turns out to be the most important, as a consequence of the quadratic dependence of  dipolar strength on domain size. 

Structural and dynamical properties of monolayers are mainly determined by inter--domain interactions and hence, for systems where the dipolar interaction is dominant, are expected to be strongly affected by the presence of domains of different sizes.

To the best of our knowledge, the effects of size polydispersity on the structural and dynamical properties of lipid monolayers have not been studied. However, for the determination of physical parameters of the  constituting lipids, size distribution has been taken into account.
Mulder~\cite{mulder03} studied the use of size distributions of circular domains in Langmuir monolayers for determining physical parameters of surfactants. He approximates the exact size distribution by a Gaussian distribution and uses a simplified theoretical analysis, in which the inter--domain interactions are treated approximately by considering equally sized domains arranged in a regular hexagonal array.
Lee \textit{et al.}~\cite{lee11} obtain the excess dipolar density by fitting the size distribution with an equilibrium thermodynamic expression. Their scheme assumes no interactions between domains, and hence it is valid for sufficiently diluted systems.

Size distribution effects, on the other hand, have long been studied in colloidal systems, mainly 
for static properties~\cite{loewen06, loewen09,bitsanis10}, 
phase transitions~\cite{bitsanis09, fasolo03,linden13,pol08,sollich11},
crystallization~\cite{tretiakov12,iacopini09,sear98,pusey87,mcrae88,bolhuis96,auer01,loewen13},
glass transition~\cite{zaccarelli15,voigtmann03,heckendorf17},
self--assembly~\cite{cabane16},
drying of colloidal dispersions~\cite{boulogne16},
determination of effective interactions~\cite{frydel05, pond11, pangburn05, pangburn06}
and dynamical properties~\cite{bitsanis10}.
In particular, for quasi--two--dimensional systems, most of the works consider binary mixture~\cite{loewen06, loewen09, loewen13}, while a few explore polydisperse colloidal suspensions~\cite{pangburn05, pangburn06, frydel05, tretiakov12}.
These study the influence of polydispersity in the determination of an effective interaction potential, when the polydisperse system is regarded as a monodisperse system.

In a previous work~\cite{rufeil16}, we proposed a novel way of estimating dipolar repulsion, using a passive method that involves the analysis of images of the monolayer with phase coexistence. 
The method is based on comparing the pair correlation function obtained from experiments with that obtained from simulations of systems of monodisperse domains interacting by a dipolar density pair potential. 
We also studied the point dipole approximation for the dipolar density pair potential, and determined an effective point dipole interaction strength that reproduces the structural properties of a system with dipolar density pair potential.

In this work, we use Brownian dynamics simulations to investigate the effects of polydispersity on the structure and dynamics of lipid monolayer models, with parameters chosen within a range of experimental interest.
For this purpose, we study the pair correlation function, the average mean square displacement and the average time--dependent self--diffusion coefficient, as a function of polydispersity and dipolar interaction strength.
In addition, we study how polydispersity would affect determination of dipolar repulsion strength from experimental data, when the method proposed in Ref.~\cite{rufeil16} is used based on a monodisperse model.

\section{Model and Theory}

\subsection{Polydisperse domain interactions}

We consider a monolayer in its two--phase liquid--condensed (LC) and liquid--expanded (LE) coexistence region, where the LC phase forms domains in the LE phase, which occupies the larger area of the monolayer. 
Because of the difference in surface densities, the LC domains possess an excess dipole density, $\sigma$, with respect to the surrounding LE phase \cite{mcconnell}. 
This originates dipolar repulsive interactions between the domains.
We model the mixed monolayer as a 2D--dispersion of domains which interact through a dipolar pair potential. Considering only dipole components perpendicularly oriented to the interface, and using the approximation of a point dipole in the center of each domain with an excluded area,
the resulting pair potential between domain $i$ and domain $j$ can be described by:
\begin{equation}
\label{U}
 u_{i,j}(r)= u_{hc}(r)+  u_{d}(r),
\end{equation}
where $u_{hc}(r)$ is a hard core repulsive potential, and
\begin{equation}
\label{Up}
 u_{d}(r)= \frac{\mu_i \mu_j}{4\pi\epsilon_0\epsilon^*}\frac{1}{r^3},
\end{equation}
where $\mu_i$ is the dipole moment of the domain $i$ representing the dipole density $\sigma$ over the domain area $A_i$, $\mu_i=\sigma A_i$, 
$r$ center--to--center domain distance, 
$\epsilon_0$ is the vacuum permittivity and $\epsilon^*$ is an effective permittivity~\cite{urbakh93} that considers the relative permittivities of the membrane, $\epsilon_m$, water, $\epsilon_w$, and air, $\epsilon_a$:
\begin{equation}
\label{epss}
 \epsilon^*=\frac{\epsilon_m^2(\epsilon_w+\epsilon_a)}{2\epsilon_w\epsilon_a}.
\end{equation}
We consider polydisperse circular domains of radius $R_i$, accordingly Eq.~\ref{Up} becomes:
\begin{equation}
\label{Upp}
 u_{d}(r)= \frac{\sigma^2}{4\pi\epsilon_0\epsilon^*}\pi^2 R_i^2 R_j^2\frac{1}{r^3}.
\end{equation}
This equation makes explicit how domain size distribution leads to interaction polydispersity.

For convenience, we define a dimensionless interaction strength:
\begin{equation}
\label{f0}
 f= \frac{\sigma ^2}{4\pi\epsilon_0\epsilon^*}\frac{R_m}{k_B T} \;,
\end{equation}\noindent
where $k_B$ is the Boltzmann's constant, $T$ the absolute temperature and $R_m$ the mean value of the radii distribution.
Then, the dimensionless dipolar pair potential takes the form:
\begin{equation}
\label{Uppp}
 \frac{u_{d}(r)}{k_B T}= f  \frac{\pi^2 R_i^2 R_j^2}{R_m^4}  \left( \frac{R_m}{r}\right)^3 .
\end{equation}

\subsection{Domain size distribution}

Most experiments on lipid monolayers present domain size polydispersity~\cite{hu06, lee11, wilke14, rufeil16, min17}. The functional form of the domain size distribution is clearly crucial to the structural and dynamical properties of the monolayer. Here, we approximate the size polydispersity by a truncated normal distribution function,
\begin{equation}
\label{gaussiana}
 P(R)  = \frac{a}{\sqrt{2\pi}\Sigma}\: \mathrm{e}^{-\frac{1}{2}\left( \frac{R-R_m}{\Sigma}\right)^2},\quad \quad 0<R<2 R_m,
\end{equation}\noindent
where $R_m$ is the mean or expectation of the distribution, $\Sigma$ is the standard deviation,
and $a$ is a normalization constant which arises from simetrically truncating the normal distribution function to exclude negative domain radii. 
The system polydispersity, $\omega$, is characterized by the ratio of the width of the distribution to its mean; $\omega=\Sigma/R_m$. 

In our simulations we use a discrete counterpart of $P(R)$ (see Fig.~\ref{distribution}), where we chose a number of domain species $N_{pol}$, and a bin width $s$, such that the radii distribution of the system is described by the set $\{R_{\alpha}, N_{\alpha}/N\}$, $\alpha=1,N_{pol}$, where $N_{\alpha}$ is the number of domains of type $\alpha$ and $N$ is the total number of domains.  
For the system with the largest polydispersity studied, $\omega=0.4$, $1\%$ of the domain radii fall outside the histograms. 

\subsection{Radial distribution function}

A key quantity for characterizing the structure of the monolayer is the radial distribution function (RDF), $g(r)$.
Considering a distribution of domains in the monolayer plane, $g(r)$ is related to the probability of finding a domain at a distance $r$ from another domain chosen as a reference point:
\begin{equation}
\label{gr}
g(r)=\frac{1}{N^2} \sum_{\substack{\alpha=1}}^{N_{pol}} \sum_{\substack{\beta=1}}^{N_{pol}} N_{\alpha} N_{\beta}g_{\alpha,\beta}(r) \; ,
\end{equation}
where $g_{\alpha,\beta}(r)$ are the partial radial distribution functions, defined as:
\begin{equation}
\label{grab}
g_{\alpha,\beta}(r)=\frac{A}{N_{\alpha} N_{\beta}} \left\langle \sum_{\substack{i=1}}^{N_{\alpha}} \sum_{\substack{j=1\\j\neq i}}^{N_{\beta}} \delta(\vec{r}-\vec{r}_i+\vec{r}_j) \right\rangle \; ,
\end{equation}
with $A=L^2$ the total monolayer area and the angular brackets indicating an equilibrium ensemble average. 
Note that $\rho_{\beta} g_{\alpha,\beta}(r)$ is the probability density of finding a $\beta$ particle at a distance $r$ from an $\alpha$ particle, where $\rho_{\beta}=N_\beta/A$ is the number density of domains with radius $\beta$. 

\subsection{Diffusion}

In order to evaluate the effects of polydispersity on the system dynamics, we studied the mean square displacement (MSD) and self--diffusion of domains.
The mean square displacement of a domain of type $\alpha$ with its center at position $\vec{r}_{1,\alpha}(t)$ at time $t$ is given by:
\begin{equation}
\label{Wa}
W_{\alpha}(t)=\frac{1}{4}\left\langle \left[ \vec{r}_{1,\alpha}(t)- \vec{r}_{1,\alpha}(0) \right]^2  \right\rangle, 
\end{equation}
where the angular brackets indicate an equilibrium ensemble average. However, many experimental results do not differentiate between domain radii. 
Hence, we evaluate the average MSD as a representative quantity:
\begin{equation}
\label{Wp}
W(t)=\sum_{\substack{\alpha=1}}^{N_{pol}}x_{\alpha}W_{\alpha}(t),
\end{equation}
where $x_{\alpha}=N_{\alpha}/N$ is the molar fraction of domains of type $\alpha$. 

The time--dependent self--diffusion coefficient, $D_{\alpha}(t)$, is defined as the time derivative of $W_{\alpha}(t)$. The short--time limit of $D_{\alpha}(t)$ corresponds to the free domain diffusion coefficient, which in this work is approximated by the diffusion coefficient of a disk in a two--dimensional simple fluid:
\begin{equation}
\label{Da0}
D_{\alpha}^0=\dfrac{k_B T}{4\pi \eta R_{\alpha}}
\end{equation}
where $\eta$ is the viscosity of the fluid. It is important to note that, in our simulations, only the ratios between the different $D_{\alpha}^0$ are relevant.

The long--time self--diffusion coefficient is defined as:
\begin{equation}
\label{DaL}
D_{\alpha}^L=\lim_{t\to\infty}D_{\alpha}(t).
\end{equation}
Analogously, the average time--dependent self--diffusion coefficient, $D(t)$, is obtained from the time derivative of $W(t)$, and the corresponding limits are:
\begin{equation}
\label{D0}
D^0=\sum_{\substack{\alpha=1}}^{N_{pol}}x_{\alpha}D_{\alpha}^0,
\end{equation}
\begin{equation}
\label{DL}
D^L=\sum_{\substack{\alpha=1}}^{N_{pol}}x_{\alpha}D_{\alpha}^L=\lim_{t\to\infty}D(t).
\end{equation}
%

\subsection{Simulations}

%
\begin{figure}[t]
\centerline{
\includegraphics[width=0.8\columnwidth]{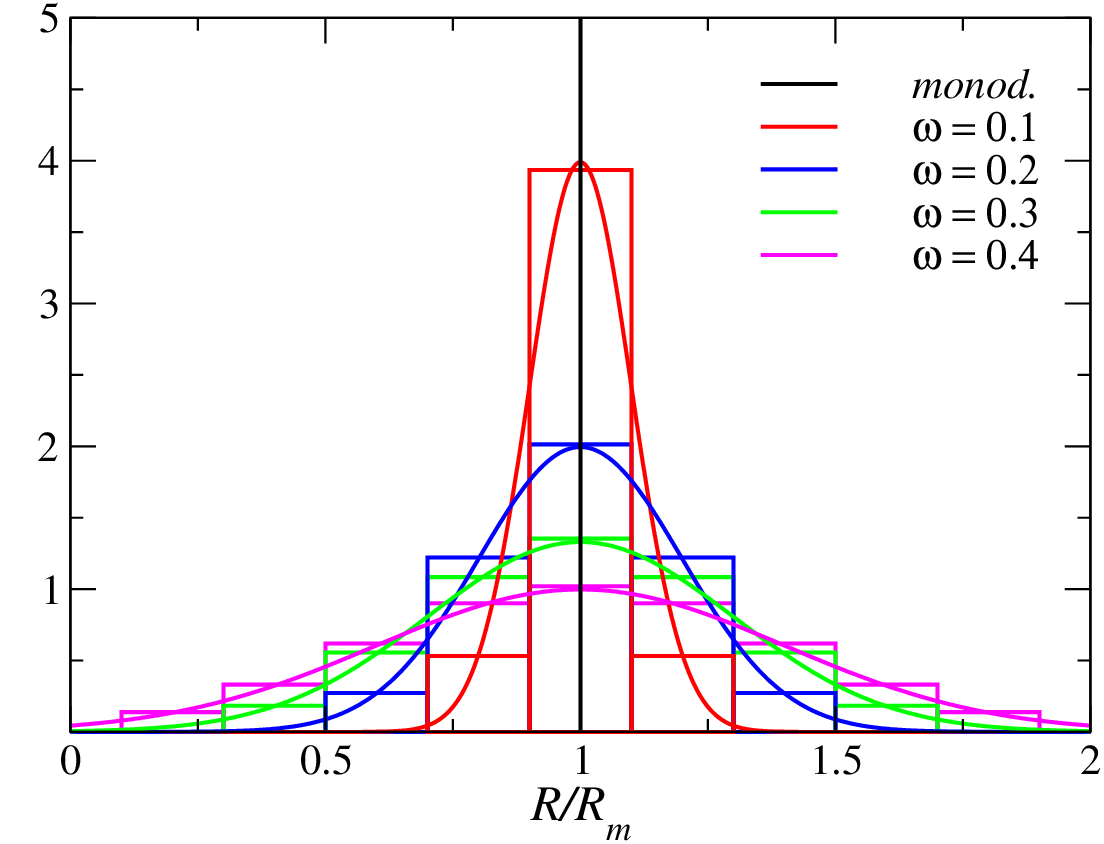} }
\caption{Size distributions used for the polydisperse systems (histograms) and their Gaussian counterpart.}
\label{distribution}
\end{figure}

We model the mixed monolayer as a two--dimensional Brownian suspension of interacting hard disks with polydisperse radii immersed in an effective fluid, each disk representing an idealized lipid domain. 
The inter--domain interactions are described by the point dipole pair interaction Eq.~(\ref{Uppp}) plus a hard disk repulsive part. 
Hydrodynamics interactions are disregarded.
To study the static and dynamical properties of the mixed monolayer model, we performed Brownian dynamics (BD) simulations.
In this scheme, the finite difference equation describing the in-plane displacement of $N$ Brownian disks immersed in a fluid during the time step $\Delta t$ is given by~\cite{ermak}
\begin{equation}
\label{bd}
 \vec{r}_{\alpha,i}(t+\Delta t)- \vec{r}_{\alpha,i}(t)=\sum_{\beta=1}^{N_{pol}} \frac{D_{\beta}^0}{k_B T} \sum_{j=1}^{N_{\beta}}  F_{\beta,j}^P \Delta t + \vec{X}_{\alpha,i}  \; ,
\end{equation}\noindent
where $\vec{r}_{\alpha,i}(t)$ is the position of domain $i$ of type $\alpha$ at time $t$, $F_{\beta,j}^P$ is the force on disk $i$ of type $\alpha$ due to the disk $j$ of type $\beta$
and $\vec{X}_{\alpha,i}$ a random displacement vector of domain $i$ of type $\alpha$ originating from solvent particle collisions. $\vec{X}_{\alpha,i}$ is sampled from a Gaussian distribution with zero mean and covariance matrix:
\begin{equation}
\left\langle \vec{X}_{\alpha,i} \right\rangle =0; \quad  \left\langle \vec{X}_{\alpha,i} \vec{X}_{\beta,j} \right\rangle=2 D_{\alpha}^0  \mathbf{I}\, \delta_{\alpha,\beta}\, \delta_{i,j}\, \Delta t,
\end{equation}\noindent
where $\mathbf{I}$ is the identity matrix, and $\delta_{i,j}$ the Kronecker delta.

The simulated systems consisted of $N$ disks with radii distribution $\{R_{\alpha},N_{\alpha}/N\}$ under periodic boundary conditions, using the minimum image convention.

The size of the simulation box, $L$, was determined from the condensed area fraction, $\phi$, defined as:
\begin{equation}
\label{phi}
\phi=\sum_{\alpha} \phi_{\alpha}; \quad \phi_{\alpha}=\frac{N_{\alpha} \pi R^2_{\alpha}}{L^2}.
\end{equation}

The system is completely characterized by the following parameters: the interaction strength $f$ (or equivalently, the dipolar density $\sigma$), the size distribution $\{R_{\alpha},N_{\alpha}/N\}$ and the total area fraction $\phi$.

Throughout this work, the area fraction is fixed at the value $\phi=0.20$, which was chosen as a typical value of experimental monolayer micrographs.

In Figure~\ref{distribution}, we show the distributions used in our studies and, in Table S1 of the electronic supplementary information (ESI), the size distribution $\{R_{\alpha},N_{\alpha}/N\}$ of each system studied can be found.
We have carried out simulation studies with four distinct polydispersities: $\omega=0.1,\omega=0.2, \omega=0.3$ and $\omega=0.4$. These systems are compared with simulations of perfectly monodisperse domains. 

The bin width was chosen as $s/R_m=0.2$. 
This choice leads to a small number of domain types for each polydispersity, and at the same time reproduces the distribution shape qualitatively well. 
For a typical experimental average radius of $R_m=5$ px, the bin width $s=1$ px is in the range of typical optical microscopy experimental error.
We verified that, for the systems studied, the Gaussian distribution is well--described by the selected discretization $\{R_{\alpha},N_{\alpha}/N\}$. The RDF and the average MSD do not change substantially if half of the bin width value is used.

Finally, the interaction strength $f$ was varied such that the system remains in its fluid phase and has experimental interest.

The time scale used in the simulations is $R_m^2/D_m^0$.
In our simulations, we used $\Delta t\,D_m^0/R_m^2 = 10^{-3}, 5.12\;10^{-4}, 2.16\;10^{-4}, 6.4\;10^{-5}$ and $8\;10^{-5} $ for monodisperse, $\omega=0.1,0.2,0.3$ and $0.4$, respectively.
These values were selected so that the dynamics of the smallest domain in each system is well--resolved.
For all systems we used $N=676$ domains. 
The number of domains of each type for the systems considered are specified in Table S1 of the ESI. 
For the systems studied, we verified that there is no system size dependency. 

\section{Results and Discussions}

\subsection{Effects of polydispersity on $g(r)$ and $MSD(t)$}

With the aim of analyzing the effect of polydispersity, we consider systems with four different interaction strengths, which correspond to the liquid regime for the selected area fraction ($\phi=0.2$), 
and for each interaction strength we vary the polydispersity from $\omega=0$ (monodisperse) to $0.4$.
Here, we present the results for $f=1.2$ and $4.8$, and, in the ESI, the intermediate values of $f=2.4$ and $3.6$ are shown (Fig. S1). 

\begin{figure}[h]
\centerline{
\includegraphics[width=0.9\columnwidth]{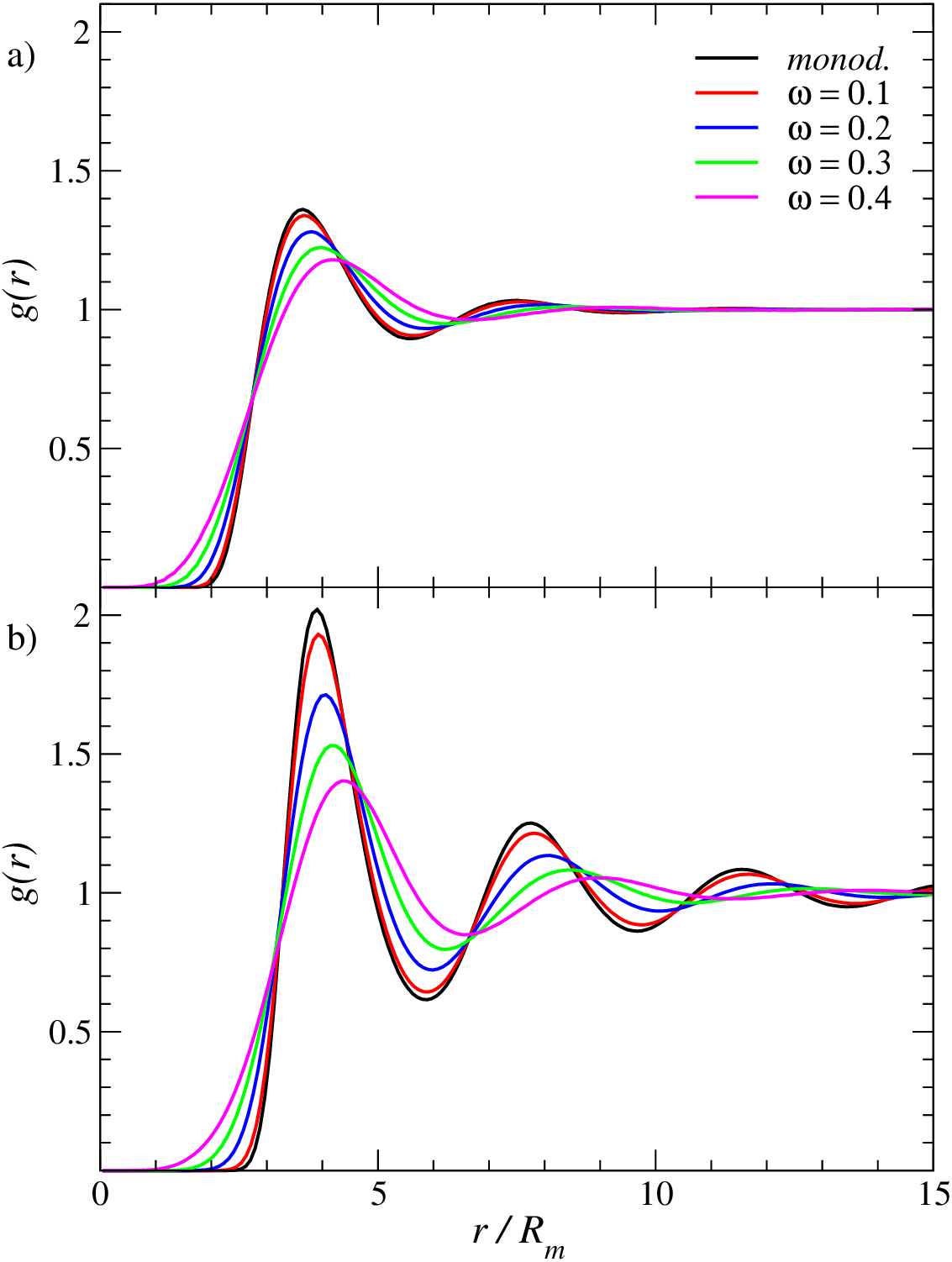}}
\caption{Radial distribution function for monodisperse and polydisperse systems with (a) $f=1.2$, and (b) $f=4.8$.}
\label{gr_f1_2_y_f4_8}
\end{figure}

Figure~\ref{gr_f1_2_y_f4_8} shows the pair correlation function for $f=1.2$ (a) and $4.8$ (b), for $\omega=0,0.1,0.2,0.3$ and $0.4$. In both cases, we observe a decrease and a broadening of the first peak as the polydispersity grows. Furthermore, there is a shift of the peak position to greater distances and the $g(r)$ start to show correlations for shorter distances. In general, polydispersity softens the peaks and minima of the $g(r)$. 
It is remarkable that, even for $\%40$ of polydispersity, there is still a well--developed first minimum and second maximum, for highly interacting systems. This can be attributed to the fact that the interaction strength grows with the fourth power of the domain size. 

\begin{figure}[h]
\centerline{
\includegraphics[width=0.9\columnwidth]{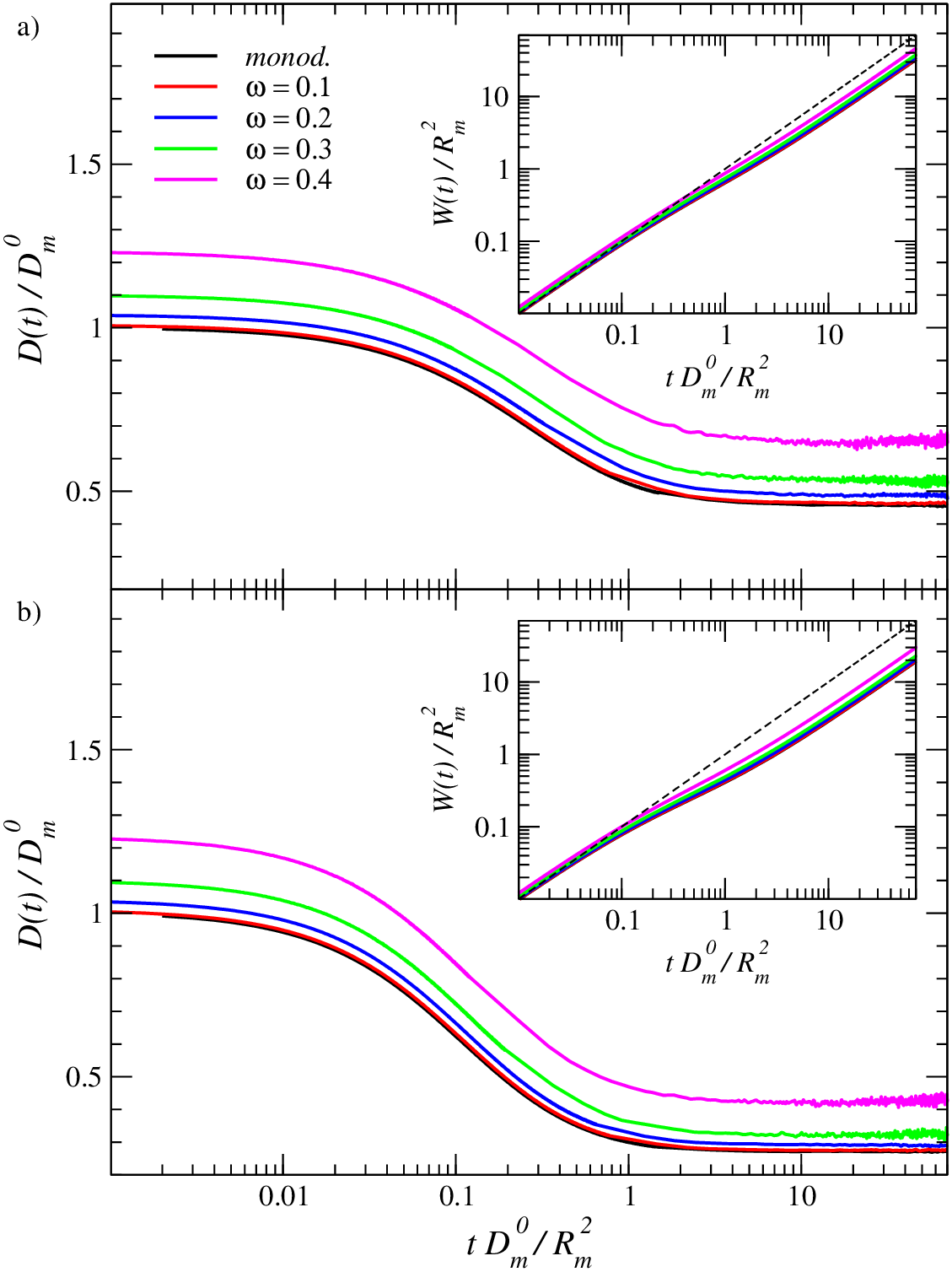}}
\caption{Average diffusion coefficient $D(t)$ and average MSD (inset) for monodisperse and polydisperse systems with (a) $f=1.2$, and (b) $f=4.8$. In the insets, the dashed line indicates the low--density limit for the monodisperse system, $W(t)=D_m^0 \;t$.}
\label{MSD_f1_2_and4_8}
\end{figure}

To analyze how polydispersity affects the dynamical properties of the monolayer, we calculated the average MSD and the average time--dependent diffusion coefficient.
Results for the selected systems are presented in Fig.~\ref{MSD_f1_2_and4_8}.
For the weak interaction $f=1.2$ (Fig.~\ref{MSD_f1_2_and4_8}a), we observe that the average short--time diffusion coefficients of polydisperse systems are larger than that of monodisperse system. This is a direct consequence of the inverse radius dependence of the short--time coefficient diffusion, $D_{\alpha}^0$, and of the symmetry of the radii distribution. Here, for the more polydisperse case, $\omega=0.4$, the difference is about $20\%$.
For the average long--time diffusion coefficients we find a similar ordering, but with slightly larger differences. In particular, for $\omega=0.4$ the difference reaches $40\%$. 
However, the relative decrease of the average long--time diffusion constant with respect to the short--time limit, $D^L/D^0$, differs by less than $15\%$. 
For all polydispersities, $D(t)/D^0$ diminish roughly $50\%$ at the long--time limit.
In the inset of Fig.~\ref{MSD_f1_2_and4_8}a, the respective average MSDs are shown. 
Note that, within the logarithmic scale, the polydisperse systems appear very similar to the monodisperse system, except for a slight translation to larger values.

Figure~\ref{MSD_f1_2_and4_8}b shows the average diffusion quantities for $f=4.8$. 
For these systems, the same qualitative behavior is observed as in previous ones. 
In particular, the short--time limits are identical
since they have identical radii distribution. 
In addition, the average long--time diffusion coefficients are smaller, because of the stronger interactions. 

Comparing the most polydisperse system, $\omega=0.4$, with the monodisperse one, it is found that the average long--time diffusion coefficient is of the order of $60\%$ larger, while the ratio $D^L/D^0$ is roughly $30\%$ larger.
In general, an enhancement of average diffusion is observed as polydispersity increases. 
This effect is stronger in the long--time regime and is also more pronounced for systems with stronger interactions.

\begin{figure}[h]
\centerline{
\includegraphics[width=0.9\columnwidth]{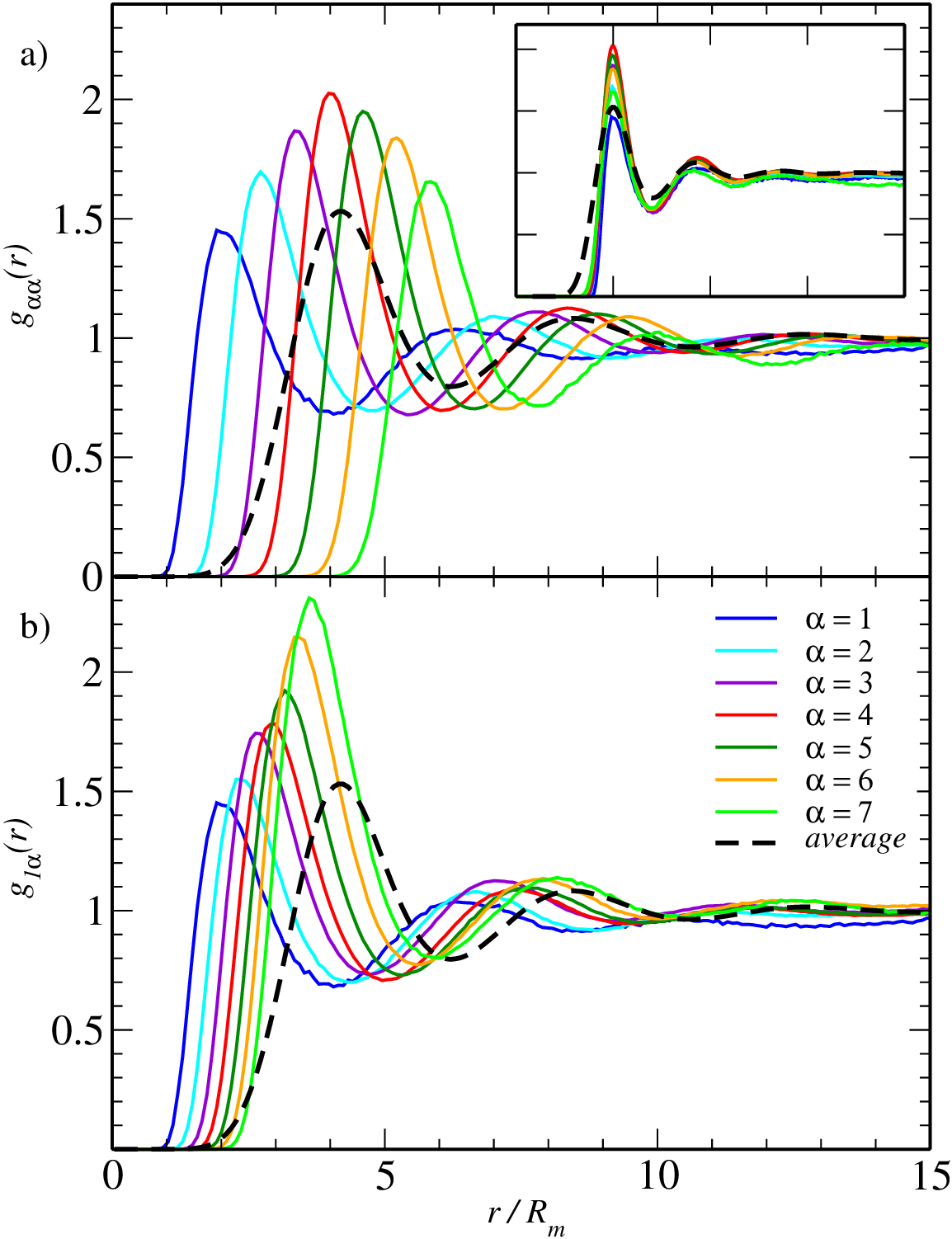}}
\caption{Radial distribution function $g(r)$ and partial RDFs $g_{\alpha,\alpha}(r)$ (a) and $g_{1,\alpha}(r)$ (b) for $\omega=0.3$ and $f=4.8$. Inset: the data have been horizontally shifted the data have been horizontally shifted to match the first peak position.}
\label{grij}
\end{figure}

To describe the structure of the polydisperse system in more detail, we show, for the system with $\omega=0.3$ and $f=4.8$, the partial radial distribution functions $g_{\alpha,\alpha}(r)$ in Fig.~\ref{grij}a and $g_{1,\alpha}(r)$ in Fig.~\ref{grij}b.
These show that the partial RDFs start to differentiate from zero at distances larger than the respective contact values, \textit{i.e.}, domains do not come into contact. 
Note that the distance between the first and the second neighbor shell and the first minimum depth are very similar for different domain sizes. This is clearly seen in the inset of Fig.~\ref{grij}a, where we have plotted $g_{\alpha,\alpha}(r)$ horizontally shifted by their respective peak positions, $r_{max}$. 
Considering the spatial correlation between the smaller domains and the other domain types, as shown in Fig.~\ref{grij}b, it is observed that
the probability of finding the smallest domains (type $\alpha=1$) around domains of other types increases with domain sizes. However, the first minimum depth of $g_{1,\alpha}(r)$ decreases with domain type $\alpha$. 
The other $g_{\alpha,\beta}(r)$, not shown here, also indicates that small domains are more probably found close to larger ones. 

\begin{figure}[h]
\centerline{
\includegraphics[width=0.8\columnwidth]{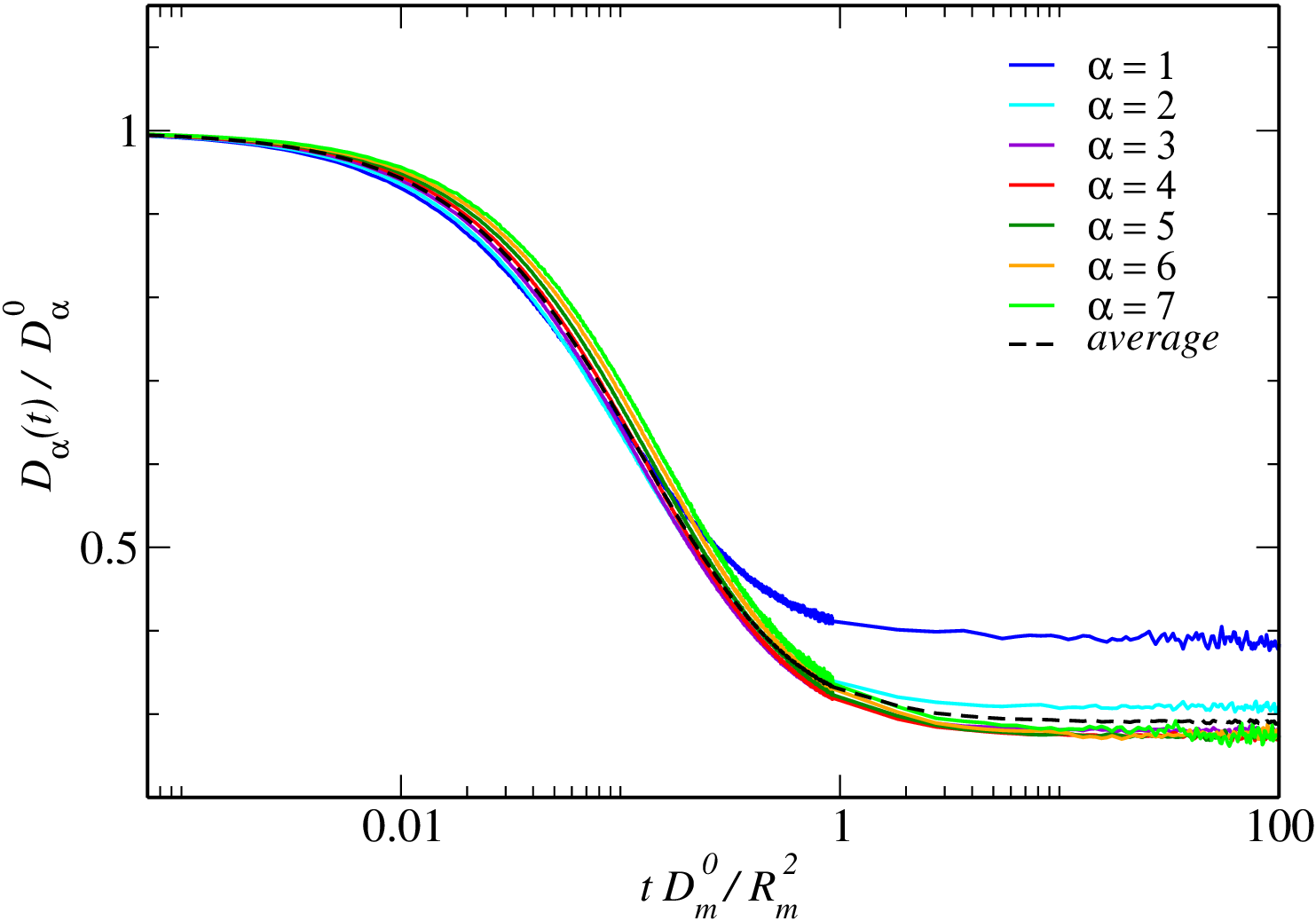}}
\caption{Normalized diffusion coefficient for each type of particle $D_\alpha (t)/D_\alpha^0$ and for the average $D(t)/D^0$ for $\omega=0.3$ and $f=4.8$.}
\label{MSDi}
\end{figure}

Focusing now on the time dependent diffusivities, in Fig.~\ref{MSDi} we show the time dependent diffusion coefficient for each domain type, $D_\alpha(t)$, and the corresponding average, $D(t)$. 
Here, to compare the relative slowdown of the dynamics between the different types, the diffusivities are normalized by their corresponding short time limit, $D^0_\alpha$.
It is observed that, except for the smallest domain type, the self--diffusion of the different domain sizes slows down similarly, reaching a long--time limit of roughly $D^L_\alpha/D^0_\alpha\simeq 0.29$. 
There is a remarkably large difference in the behavior of the smallest domains, which show a much smaller slowdown than the others, i.e., they are able to leave the neighbor cages more easily. 
This could be attributed to the fact that the dipolar strength of the domains grows quadratically with the radius.

\subsection{Monodisperse vs. polydisperse models}

In lipid monolayers with phase coexistence, the dipolar repulsion $\sigma$ ($f$) is usually not a known parameter. 
One method to determine it is to fit the experimentally measured RDF with one obtained from simulations, using $\sigma$ as the only adjustable parameter\cite{rufeil16}. 
This assumes monodisperse distribution of domain radii, and hence monodisperse inter--domain interactions.
Therefore, it is important to assess the effects of polydispersity on the determination of $\sigma$ using this method.
For this purpose, we used the $g(r)$ from the monodisperse systems studied in this work as the reference RDF, as if they were previously fitted to experimental data sets.
Then, we fitted the reference $g(r)$ with the polydisperse systems shown in Fig.~\ref{distribution} (as was already stated $\phi=0.2$ for all systems) and we analyzed how the interaction strength varies with $\omega$. 
Note that, as in our previous work~\cite{rufeil16}, only the first peak height of the reference $g(r)$ is considered in the fitting procedure.

\begin{figure}[h]
\centerline{
\includegraphics[width=0.9\columnwidth]{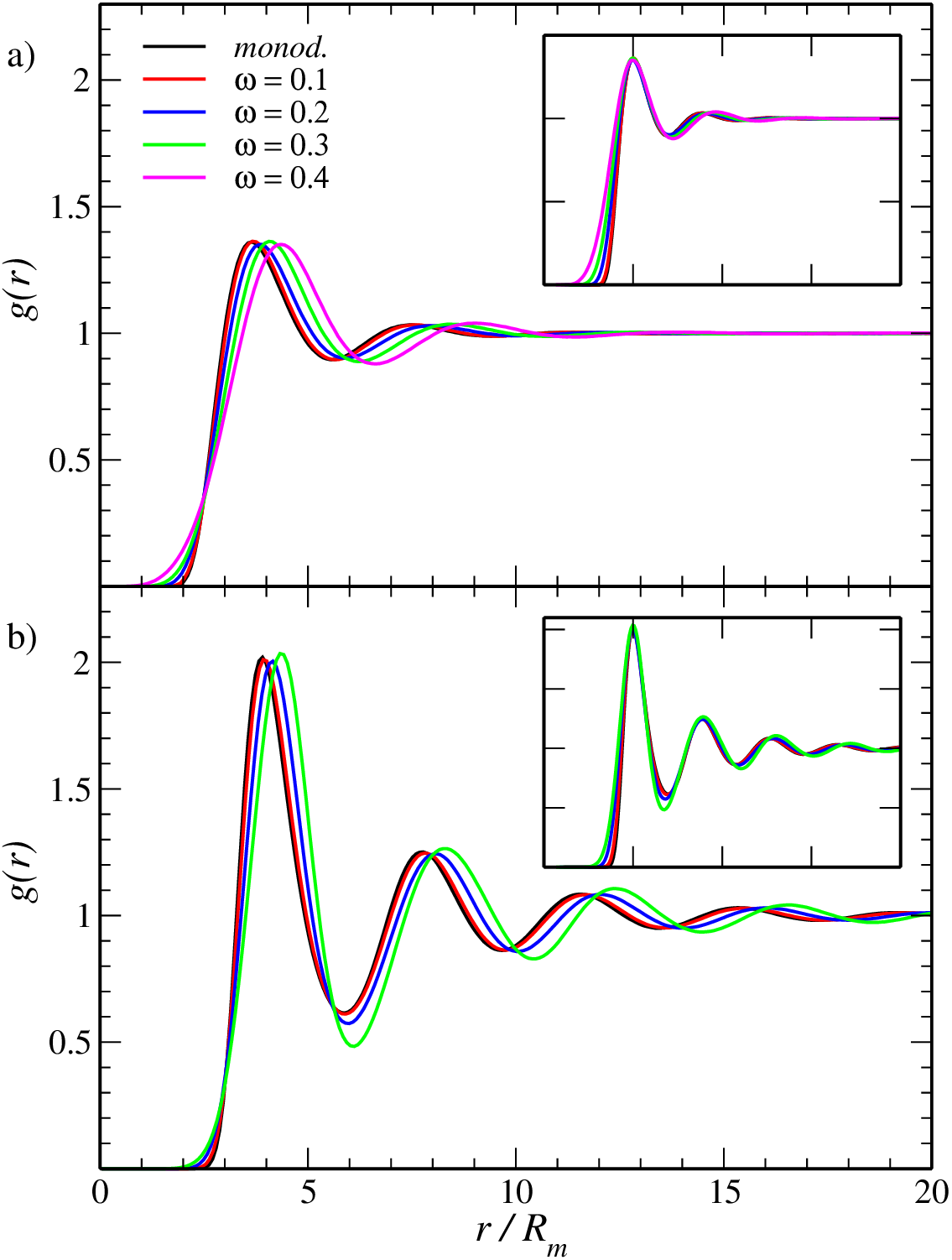}}
\caption{Radial distribution function for systems with (a) $g(r_{max})=1.35$ and $f_m=1.2$ (monodisperse), $f=1.3$ ($\omega=0.1$), $f=1.6$ ($\omega=0.2$), $f=2.4$ ($\omega=0.3$), and , $f=3.6$ ($\omega=0.4$), and (b) $g(r_{max})=2.01$ and $f_m=4.8$ (monodisperse), $f=5.5$ ($\omega=0.1$), $f=9.5$ ($\omega=0.2$) and $f=25$ ($\omega=0.3$).
Inset: the data have been horizontally shifted to match the first peak position.}
\label{gr_fit_1_2_y_4_8}
\end{figure}

Figure~\ref{gr_fit_1_2_y_4_8} shows the results for systems with $g(r_{max})=1.35$ (a) and $g(r_{max})=2.01$ (b) (monodisperse models with $f_m=1.2$ and $4.8$, respectively). In the ESI, systems with $g(r_{max})=1.63$ and $g(r_{max})=1.84$ ($f_m=2.4$ and $3.6$, respectively) are shown (Fig. S3). 

The RDF from the monodisperse system with $g(r_{max})=1.35$ (Fig.~\ref{gr_fit_1_2_y_4_8}a) is qualitatively well captured by the polydisperse systems, except for short distances, where for more polydisperse systems the $g(r)$ start to grow at shorter distances, as expected. 
In the inset, we have plotted the $g(r)$ horizontally shifted so that the first peak position coincides. This clearly shows that the distance between the different neighbor shells and the depth of minima are very similar for all polydispersities. This agreement is striking, since we are comparing systems with polydispersities as large as $40\%$.
 
For a more structured system with $g(r_{max})=2.01$, shown in Figure~\ref{gr_fit_1_2_y_4_8}b, we observe similar behavior. However, for polydispersities larger than $20\%$, the depth of the first minimum starts to differentiate, at the same time as the third peak position begins to dephase. This can be seen in the inset. 
Note that, for $40\%$ polydispersity, the fit is already not possible, i.e., there is no interaction strength for which the resulting $g(r)$ reaches the maximum value of $2.01$. 

\begin{figure}[h]
\centerline{
\includegraphics[width=0.8\columnwidth]{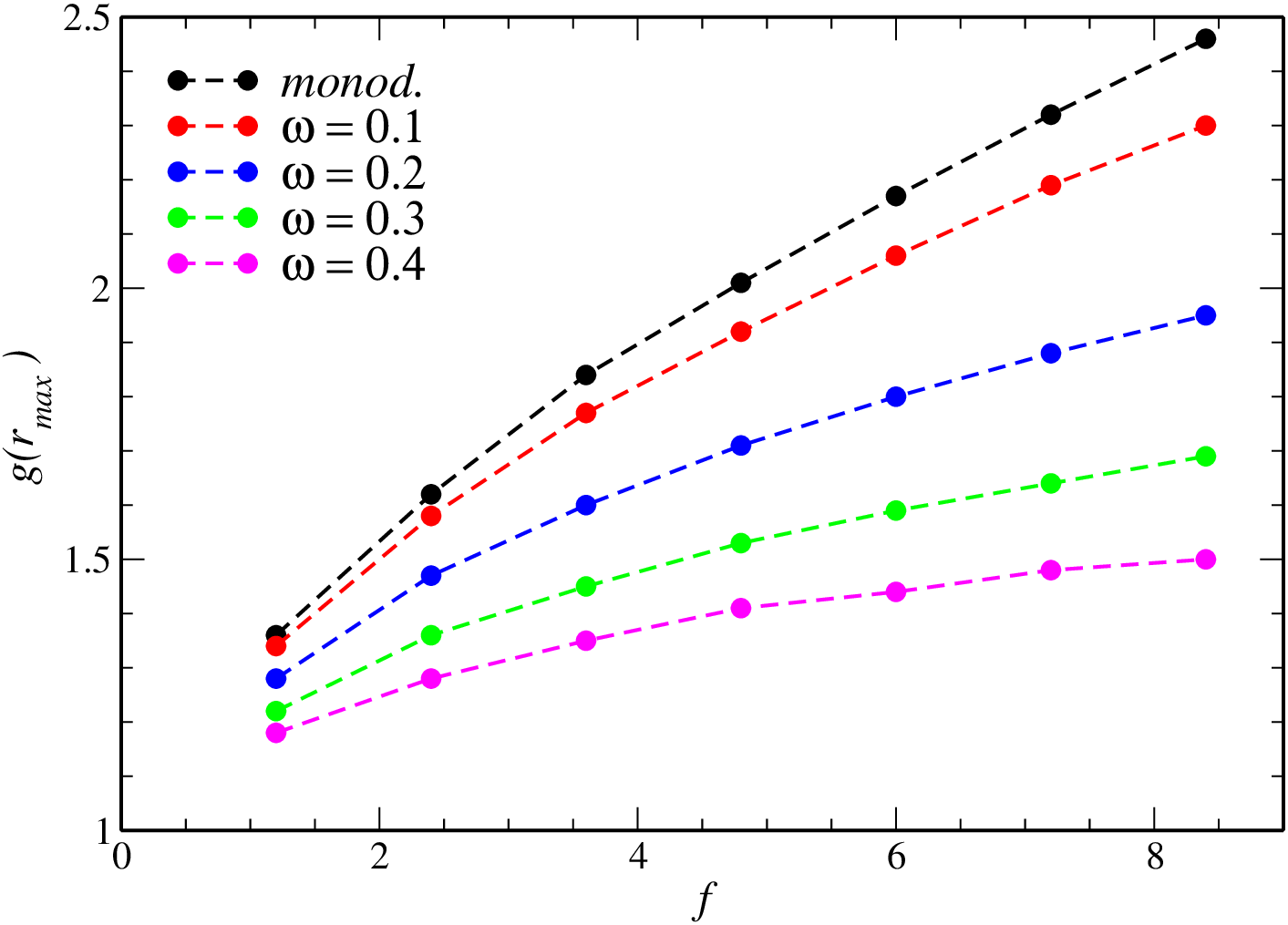} }
\caption{Maximum of $g(r)$ as a function of the interaction strength for monodisperse and polydisperse models. The lines are a guide for the eye.}
\label{fo_vs_grmax}
\end{figure}

In general, to reach a certain value of $g(r_{max})$ a larger interaction strength is needed as the polydispersities increase. However, not all the values of $g(r_{max})$ can be obtained for systems with large polydispersity.
Figure~\ref{fo_vs_grmax} shows $g(r_{max})$ as a function of the interaction strength $f$ for the different polydispersities studied here.
It can be seen that the values $g(r_{max})$ tend to saturate for high interaction strength.

For experimental systems with $\phi=0.20$ and approximately Gaussian size distributions, Figure~\ref{fo_vs_grmax} can be used as a working curve to estimate $f$ (or $\sigma$) directly from the experimental $g(r)$ without implementing any simulation, generalizing the method presented in Ref.~\cite{rufeil16} to polydisperse systems.

At this point, we introduce a new parameter $ \Gamma $, which combines the number density, $\rho = N/A$, and the interaction strength, $f$, in one independent dimensionless parameter. Namely,
\begin{equation}
\label{Gamma}
 \Gamma =  f \pi^2\frac{R_m^3}{r_m^3} , 
\end{equation}
where $r_m=\rho^{1/2}$ is the mean geometrical distance between domains.

Systems for which the hard disk interactions can 
be disregarded (i.e., strongly interacting or low density systems) are
completely determined by $ \Gamma $, the scaled domain radii of the different species, $\lambda_{\alpha} = R_{\alpha}/R_m$, and their corresponding molar fractions, $x_{\alpha}=N_{\alpha}/N$.
Note that the parameter space $\{f,\phi,R_{\alpha},x_{\alpha}\}$ is mapped to $\{\Gamma,\lambda_{\alpha},x_{\alpha}\}$.

\begin{figure}[h]
\centerline{
\includegraphics[width=0.8\columnwidth]{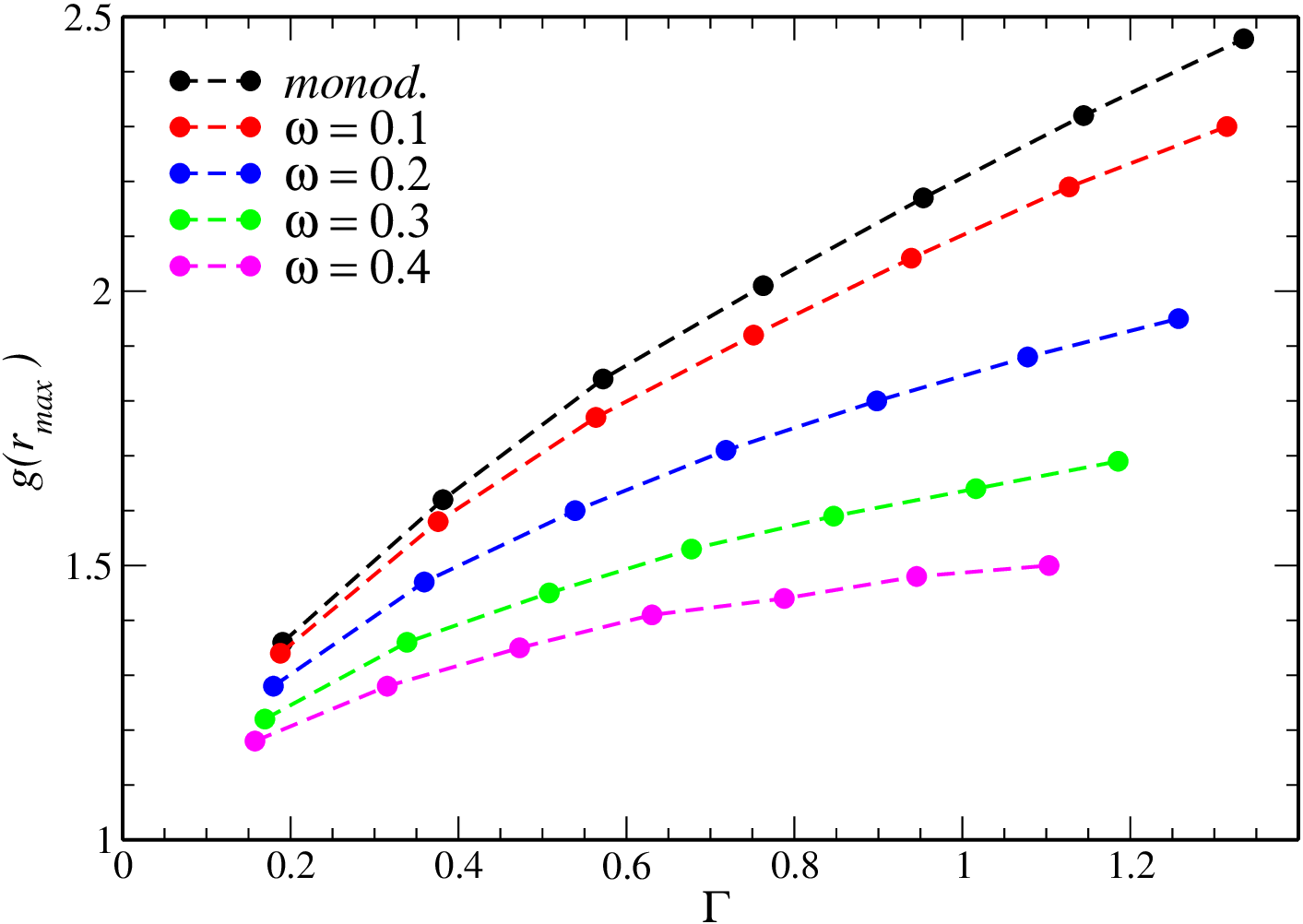} }
\caption{Maximum of $g(r)$ as a function of $\Gamma$ for monodisperse and polydisperse systems. The lines are a guide for the eye.}
\label{G_vs_grmax}
\end{figure}

Given that, for the systems studied here, the hard disk interaction turned out to be irrelevant, Fig.~\ref{fo_vs_grmax} may be recast to show the
maximum of the RDF as a function of the $\Gamma$, as shown in Fig.~\ref{G_vs_grmax}.
In this way, if the size polydispersity can be described by a Gaussian distribution with $\{\omega, R_m\}$, and $g(r_{max})$ and $\rho$ are determined from the experiments, the $\Gamma$ value could be estimated directly from this figure, and subsequently the dipolar density can be obtained. 

\begin{figure}[h]
\centerline{
\includegraphics[width=0.8\columnwidth]{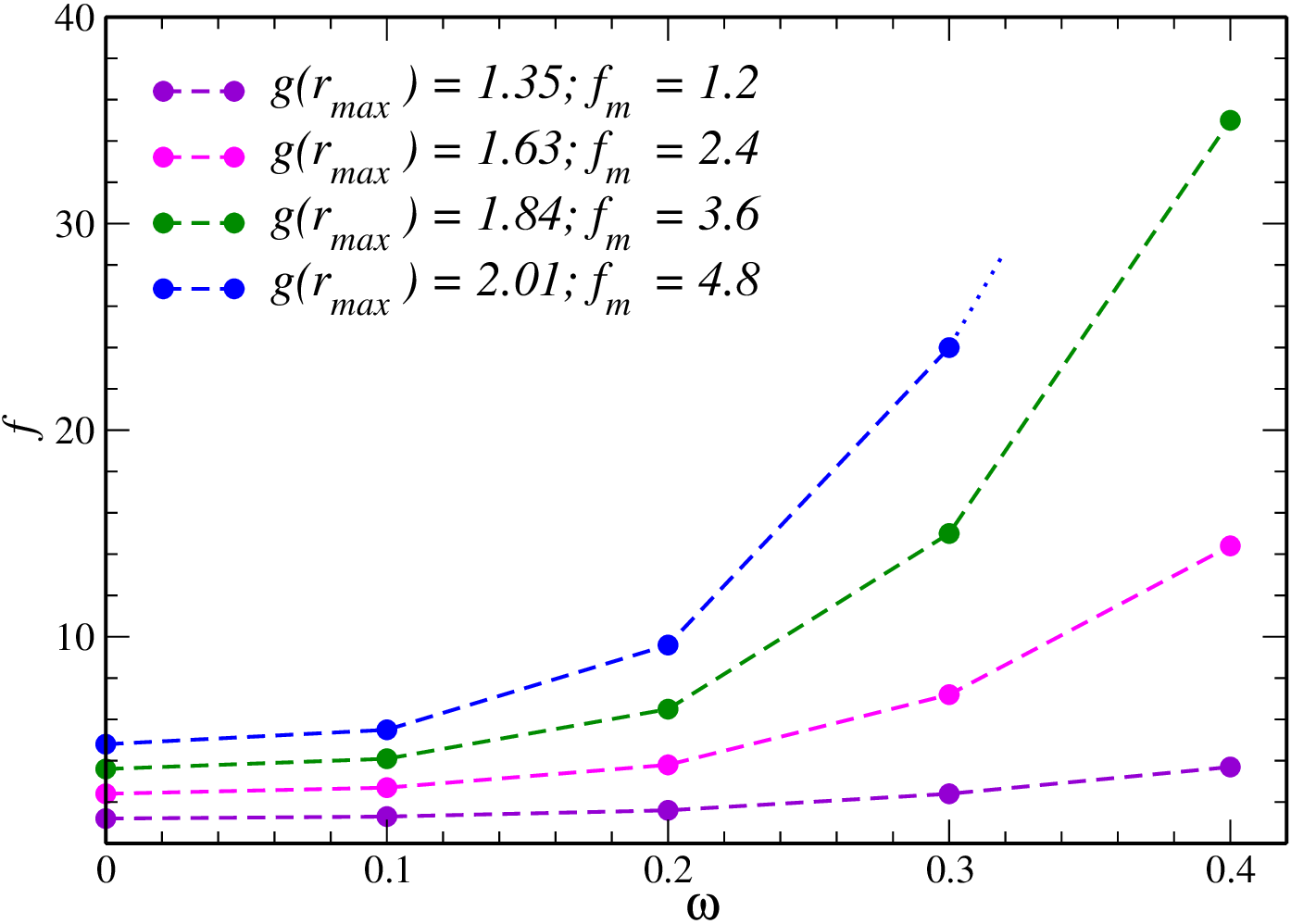} }
\caption{Interaction strength of the polydisperse models that lead to a similar structure to that of the monodisperse model, for different monodisperse systems. The lines are a guide for the eye.}
\label{esquema_fit}
\end{figure}

To further study how polydispersity affects the structure, Figure~\ref{esquema_fit} shows the interaction strength needed to obtain a certain $g(r_{max})$ as a function of $\omega$. The curves for $g(r_{max})=1.35$ and $g(r_{max})=2.01$ correspond to the RDFs shown in Fig.~\ref{gr_fit_1_2_y_4_8} a and b, respectively. 
Here, it can be clearly seen that the interaction strength for a given $g(r_{max})$ increases notably faster for more structured systems and, in particular, for $g(r_{max})=1.35$ and $\omega=0.3$, $f/f_m=2$ and for $g(r_{max})=2.01$ and $\omega=0.3$ $f/f_m=5.21$. 
The dotted line for the case $g(r_{max})=2.01$ indicates that no system with $\omega=0.4$ is able to reach this peak height.

\begin{figure}[h]
\centerline{
\includegraphics[width=0.9\columnwidth]{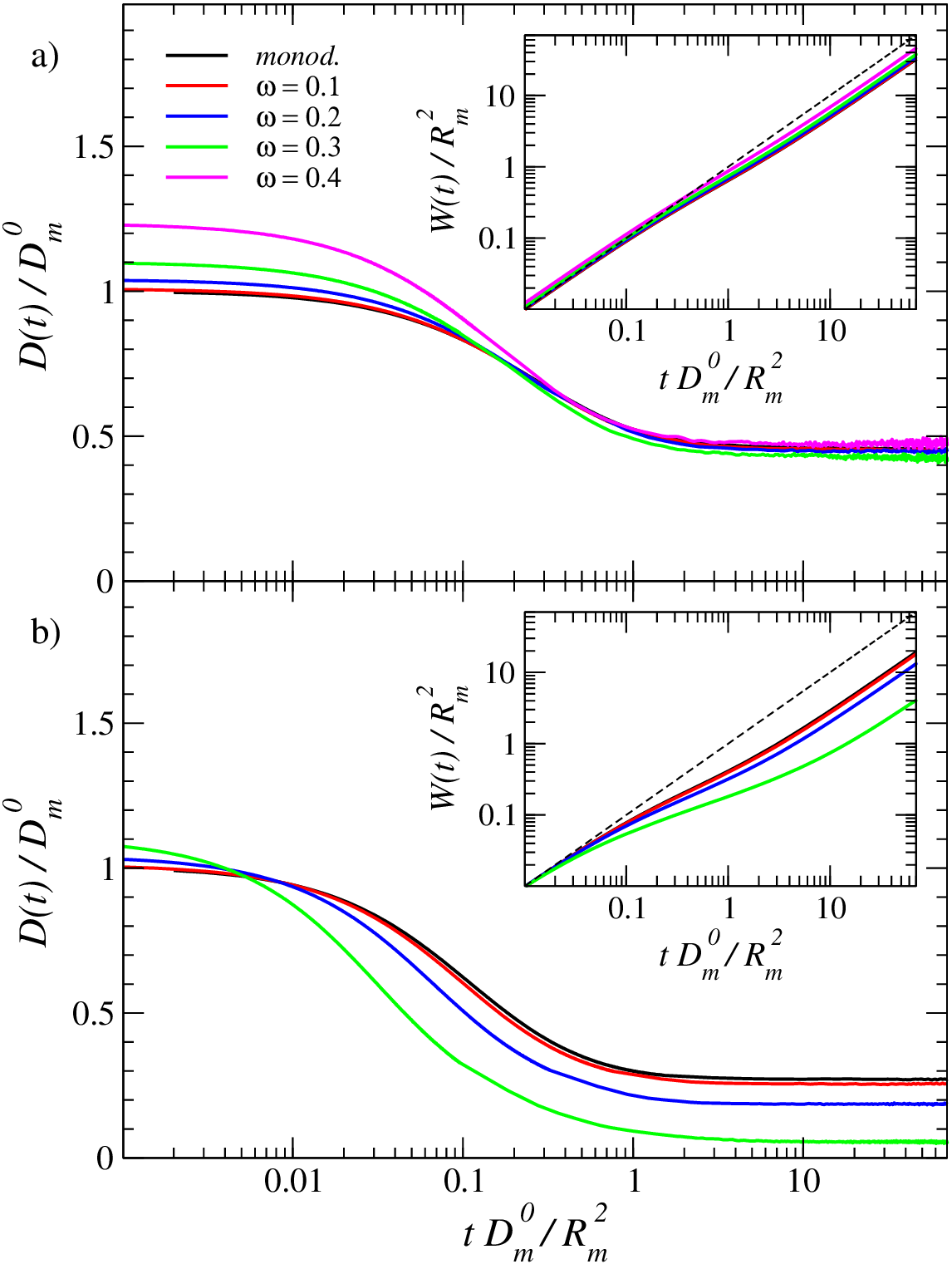}}
\caption{Average diffusion coefficient and average MSD (inset) for monodisperse and polydisperse models that lead to similar structures (see fig.~\ref{gr_fit_1_2_y_4_8}) with (a) $g(r_{max})=1.35$, $f_m=1.2$, and (b) $g(r_{max})=2.01$, $f_m=4.8$. In the insets, the dashed line indicates the low--density limit for the monodisperse system, $W(t)=D_m^0 \;t$.}
\label{MSD_fit_1_2_and_4_8}
\end{figure}

Finally, we consider the dynamics of the systems studied shown in Figure~\ref{gr_fit_1_2_y_4_8}. 
The corresponding average diffusion coefficient and MSD are shown in Figure~\ref{MSD_fit_1_2_and_4_8}.

It is remarkable that, for the less structured system (Fig.~\ref{MSD_fit_1_2_and_4_8}a), polydispersity does not much affect the intermediate and long--time average dynamical quantities.
However, a completely different scenario occurs for the more structured system (Fig.~\ref{MSD_fit_1_2_and_4_8}b). In this case, up to $\omega=0.1$, the average diffusion coefficient behaves similarly to that of the monodisperse system. On the other hand, for larger polydispersities, $D(t)$ strongly deviates from the monodisperse system. 
The system reaches the sub--diffusive regime faster, as polydispersity increases. Besides, the long--time average diffusion coefficient is smaller for larger polydispersities.
This is probably a consequence of the dependence of the interaction strength on the domain sizes, which is evident for strongly interacting systems.

%

\section{Conclusions}

In this work, we studied the influence of domain--size polydispersity on the structure and dynamics of model lipid monolayers with phase coexistence at the air--water interface.
The size--polydispersity was modeled as a discretized Gaussian distribution. 

Studying the monolayer structure, we found a decrease and a broadening of the first peak of the RDF as the polydispersity grows. Furthermore, a shift of the peak position to greater distances and the occurrence of correlations for shorter distances were observed. Notably, highly interacting systems ($f=4.8$) presented a well--developed first minimum and second maximum, even for $40\%$ of polydispersity. 
For all the systems studied, the partial RDF starts to differentiate from zero at distances larger than the respective contact values. This indicates that domains do not come into contact and hence that the hard--disk interaction is not relevant. 
Regarding the spatial correlation between the smaller domains and the other domain types, it was observed that the probability of finding the smallest domains around domains of other types increases with domain sizes. 

Analyzing the domain dynamics, we found an enhancement of the average diffusion as polydispersity increases. This is more pronounced in the long--time regime. For systems with stronger interactions, the overall enhancement is more noticeable.
It was also found that the self--diffusion of the smallest domains shows a much smaller slowdown than the other domain sizes, i.e., they are able to leave the neighbor cages more easily. 

We also studied the effects of polydispersity in the determination of $f$ (or $\sigma$) by the method proposed in Ref.~\cite{rufeil16}, where a monodisperse model is used to fit the experimental RDF. 
For the systems considered, it was found that polydispersity strongly affects the value of $f$ obtained, which is greatly underestimated if polydispersity is not considered. 
Only for the experimental system with small polydispersities and/or weak interactions is achieved a good approximation.

It is remarkable that, even for large polydispersities, the fitted RDFs result in good agreement with the reference one; they have the same second and third peak heights and neighbor shell distances. They mainly differ in the depth of the first minimum and for distances where the $g(r)$ starts to grow.

With regard to the dynamics, on the other hand, only for weak interactions or small polydispersities does the average time--dependent diffusion coefficient agree with the reference system. For stronger interactions or larger polydispersities, a noticeable slow--down is observed in the average dynamics.

Finally, the method proposed in Ref.~\cite{rufeil16} may be straightforwardly generalized to include polydispersity by directly fitting the experimental $g(r)$ with a model that accounts for the measured domain size distribution in the simulations. 
Alternatively, using a Gaussian size--distribution model and for a selected range of $\{f,\omega\}$, a set of figures, like Fig.~\ref{fo_vs_grmax}, can be generated for different values of $\phi$, for later use as working curves to estimate $f$ (or $\sigma$) directly from the experimental data without implementing any simulation.
In particular, for a system in which the hard disk interaction can be disregarded, only one working curve is needed (Fig.~\ref{G_vs_grmax}).

\subsection*{Acknowledgement}
The authors acknowledge financial support from CONICET (Consejo Nacional de Investigaciones Cient\'ificas y T\'ecnicas, Argentina), SECyT--UNC (Secretar\'ia de Ciencia y T\'ecnica de la Universidad Nacional de C\'ordoba, Argentina) and FonCyT (Fondo para la Investigaci\'on Cient\'ifica y Tecnol\'ogica, Argentina).

%

\bibliography{mr_bib} 
\bibliographystyle{rsc} 

\end{document}